\documentclass[twocolumn,prl,preprintnumbers,showpacs,amsmath,amssymb]{revtex4}
\usepackage{graphicx}
\usepackage{dcolumn}
\usepackage{bm}
\begin{document}

\title {Families of superhard crystalline carbon allotropes induced via cold-compressed graphite and nanotubes}
\author{Haiyang Niu$^1$, Xing-Qiu Chen$^1$,$^*$ Shibing Wang$^2$, Dianzhong Li$^1$, Wendy L Mao$^{2,3}$, Yiyi Li$^1$}
\affiliation{$^1$ Shenyang National Laboratory for Materials
Science, Institute of Metal Research, Chinese Academy of Sciences,
Shenyang 110016, China} \email[Corresponding author:
]{xingqiu.chen@imr.ac.cn}

\affiliation{$^2$ Department of Geological and Environmental
Sciences, Stanford University, Stanford, California 94305, USA}

\affiliation{$^3$ Stanford Institute for Materials and Energy
Science, SLAC National Accelerator Laboratory, Menlo Park,
California 94025, USA}

\date{\today}

\begin{abstract}

We report a general scheme to systematically construct two classes
of structural families of superhard \emph{sp}$^3$ carbon allotropes
of cold compressed graphite through the topological analysis of odd
5+7 or even 4+8 membered carbon rings stemmed from the stacking of
zigzag and armchair chains. Our results show that the previously
proposed M, bct-C$_4$, W and Z allotropes belong to our currently
proposed families and that depending on the topological arrangement
of the native carbon rings numerous other members are found that can
help us understand the structural phase transformation of
cold-compressed graphite and carbon nanotubes (CNTs). In particular,
we predict the existence of two simple allotropes, R- and P-carbon,
which match well the experimental X-ray diffraction patterns of
cold-compressed graphite and CNTs, respectively, display a
transparent wide-gap insulator ground state and possess a large
Vickers hardness comparable to diamond.

\end{abstract}


\maketitle

Carbon exhibits numerous allotropes (fullerenes, carbon
nanotubes(CNTs), graphene, graphite, diamond and amorphous carbon)
thanks to its ability to form $sp$-, $sp^2$- and $sp^3$-hybridized
bonds \cite{Miller97,Hirsch10,Ehrenfreund10}. It is well-known that
compression of graphite at high pressure ($>$15 GPa) and high
temperature ($>$ 1300 K)\cite{Bundy96} leads to the formation of
cubic or hexagonal diamond. In contrast, cold compression of
graphite \cite{Wendy03,
Bundy96,Aust63,Bundy67,Goncharov89,Goncharov90,Hanfland89,
Zhao89,Utsumi91,Hanfland90,Yagi92,Montgomery11}, single-walled and
multi-walled CNTs \cite{Wang2004,Tang2000,Khabashesku2002,
Popov2002,Popov2003,Bucknum2006,Kumar2007} results in superhard
allotropes of carbon, which were found to be intrinsically different
from hexagonal (or cubic) diamond \cite{Wendy03,Wang2004}. Upon
pressure release, the obtained cold compressed graphite phase can be
quenched only at low temperature ($<$ 100 K) \cite{Miller97} whereas
CNT phases compressed above 75 GPa can be quenched at room
temperature \cite{Wang2004}. In addition, these high pressure phases
exhibit superior mechanical performance with the ability to indent
single-crystal diamond \cite{Wendy03,Wang2004}, indicating at least
comparable hardness to diamond.

However, the experimental crystal structures of these cold
compressed phases of graphite and CNTs remain heavily debated. In an
effort to shed light on the puzzling structural problem of cold
compressed graphite several superhard \emph{sp}$^3$-hybridized
candidates (monoclinic M carbon \cite{Li09,Oganov06}, body-centered
tetragonal bct-C$_4$ carbon \cite{Saito05, Umemoto10}, orthorhombic
W \cite{Wang11} and Z carbon \cite{Selli,Amsler}) have been
proposed. To date, among all these proposed allotropes the most
stable one is \emph{Z} carbon \cite{Amsler}, computationally
predicted through the minima hopping method (MHM), which is exactly
the same as the \emph{oC}16-II phase proposed recently by
metadynamics simulations of structural transformations \cite{Selli}
and the C-centered orthorhombic C$_8$ phase proposed by particle
swarm optimization (CALYPSO) \cite{Zhao}. Z carbon was thought to be
the best candidate \cite{Amsler} so far that can explain the
experimental XRD peaks and Raman active mode for cold compressed
graphite \cite{Wendy03,Amsler}. Although Zhao {\em et al.}
\cite{Zhao} argued there was a matching problem with the XRD of cold
compressed graphite, they claimed \cite{Zhao} that the Z (namely,
C$_8$) phase can be interpreted as the structural solution of the
quenchable superhard carbon phase recovered from cold compressed CNT
bundles \cite{Wang2004}. However, we found that its XRD patterns
still differ significantly compared to the experimental data (as
discussed below).

If we pay more attention to the previously proposed allotropes, both
W and M carbon phases can be described as corrugated graphite sheets
interconnected by an alternating sequence of odd 5+7 membered rings
of carbon \cite{Selli}. Similarly, both bct-C$_4$ and Z carbon can
be characterized by alternating even 4+8 membered rings. Considering
all possible even and odd rings of carbon (4+8 membered rings in
bct-C$_4$ and Z carbon, 5+7 membered rings in W and M carbon,
6-membered rings in diamond), many additional possible combinations
could be expected. Therefore, it would be highly desirable to seek
\emph{a general scheme} to understand the principles of the
structural formation of possible alternative superhard allotropes
related to cold compressed graphite and CNTs.

In this paper, through first-principles calculations (for details,
see \cite{RefAdd}), we report on two S and B families of
\emph{sp}$^3$-hybridized superhard carbon allotropes by discussing
the topological stacking of zigzag and armchair carbon chains
consisting of odd 5+7 and even 4+8 membered ring patterns (see Fig.
S1 \cite{RefAdd}). Our analysis demonstrates that, after introducing
the hexagonal rings which separate the periodic 5+7 or 4+8 membered
patterns, a series of new structures can be readily created. We
further elucidate the energetic, mechanical and electronic
properties of the obtained novel phases, confirming that all these
phases are wide-gap transparent and superhard insulators. In
particular, our currently proposed P carbon as well as several other
new phases are energetically more favorable than the previously
known most stable Z carbon \cite{Selli,Zhao} over a large range of
pressures. Moreover, we confirm that R carbon of the S-family and P
carbon of the B-family are the most likely candidates so far to
match the experimental data of the cold-compressed graphite
\cite{Wendy03} and CNTs \cite{Wang2004}, respectively.

\begin{figure}[b!]
\begin{center}
\includegraphics[width=0.45\textwidth]{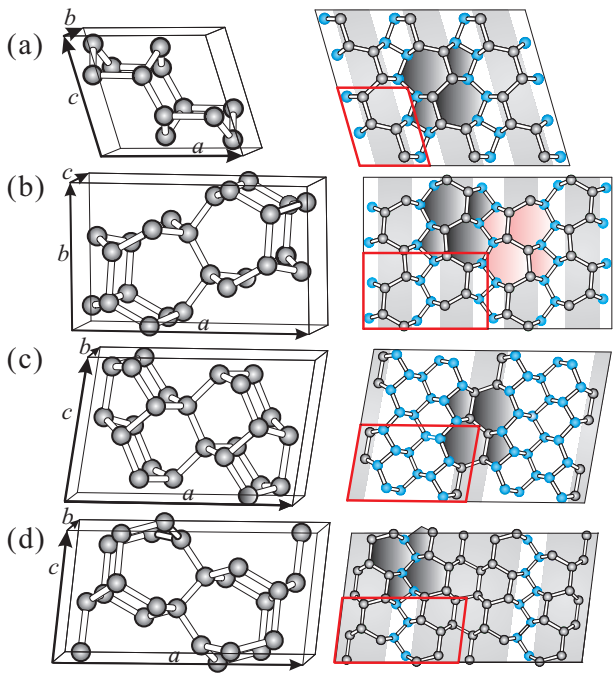}
\end{center}
\caption{(color online) (a), (b), (c) and (d) illustrate the
crystallographic structures of the proposed S carbon, R carbon,
S-S$_1$Z$_{2}$, and S-S$_1$A$_2$ phases. Left column:
three-dimensional crystalline structures and right column: the
two-dimensional (2D) projections of the 2$\times$2$\times$1
supercell along their b- or c-axis. For S carbon, the
3$\times$1$\times$2 supercell is used. The grey background
highlights the armchair chains and white background denotes the
zigzag chains.}\label{fig1}
\end{figure}

\begin{figure}[b!]
\begin{center}
\includegraphics[width=0.48\textwidth]{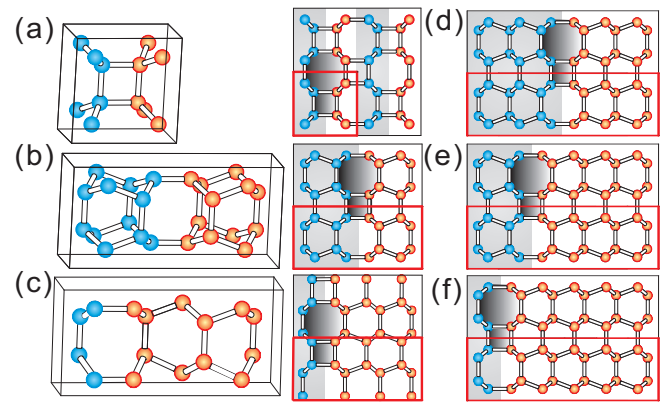}
\end{center}
\caption{(color online) (a), (b) and (c) illustrate the
crystallographic structures of the bct-C$_4$ carbon, Z carbon, P
carbon (this work). Left column: three-dimensional unit cells and
right column: the 2D projections of the 2$\times$2$\times$1
supercell of bct-C$_4$ and of the 1$\times$2$\times$1 supercells of
Z carbon along their c-axis, and the 2D projections of the
1$\times$1$\times$2 supercells of P carbon along its b-axis. (d),
(e) and (f) show the 2D-projections in the 1$\times$2$\times$1
supercells along the c-axis for three typical lattice structures of
the B-B$_1$A$_4$ series.}\label{fig2}
\end{figure}

Each graphite sheet (i.e., graphene) can be described as an ordered
array consisting of infinite long zigzag or armchair chains of
carbon atoms. The direct connection between zigzag and armchair
chains results in a tilt grain boundary \cite{Grantab} which is
interconnected by 5-membered and 7-membered rings of carbon,
different form the ideal 6-membered ring in graphene. Similarly, the
two-dimensional (2D) projections along the \emph{c}-axis for M
carbon and W carbon of cold-compressed graphite \cite{Amsler} are
also composed of 5+7 ring patterns, realized by the repeated
stacking of zigzag and armchair chains of carbon in Fig. S2
\cite{RefAdd}. The most important feature of both M and W carbon is
that each armchair chain or each zigzag chain is not infinitely
long. It is recognized that carbon bonds are combined in group of
four armchair or zigzag bonded chains. As illustrated in Fig. S2
\cite{RefAdd}, these four-bonds chains are interconnected by a step.
Interestingly, if the directly connected zigzag and armchair chains
in M carbon would be infinitely long (without any step), the simple
S carbon phase could be easily realized. As shown in Fig. 1a, its
2D-projection can be described as the repeating parallel array of
directly connected zigzag and armchair chains, consisting of
diagonally opposed 5+7 ring pattern. If this 5+7 ring pattern is
further twofold rotated, another new structure called R carbon can
be formed (Fig. 1b). Therefore, the unit cell of R carbon is twice
the size of S carbon.

The structures of S carbon and R carbon can be further modified. As
shown in Fig. 1c and 1d, by inserting two zigzag or armchair chains
to separate the 5+7 ring pattern in S carbon, two new structures
(S-S$_1$Z$_2$ and S-S$_1$A$_2$) can be realized. Here, the notation
of S come from the 5+7 ring patterns in S carbon (S$_1$ means only
one S unit), whereas Z and A denote the infinitely long zigzag and
armchair chains (Z$_2$ and A$_2$ refers to two zigzag and armchair
chains), respectively. Furthermore, it is possible to continuously
extend the number of Z and A chains to construct a series of new
structures named S-S$_m$Z$_{2n}$ and S-S$_m$A$_{2n}$. Here, the
parameters \emph{m} and \emph{n} are integers (0, 1, 2
$\cdot$$\cdot$$\cdot$). Following this nomenclature, R carbon can be
described as S-S$^{\prime}_2$ in which S$^\prime$ denotes the
existence of the twofold rotated symmetry of 5+7 ring pattern at
variance from the normal 5+7 ring in S carbon. Following similar
considerations, S$^\prime$ can be further extended into a series of
new structures called S-S$^\prime_m$Z$_{2n}$ and
S-S$^\prime_m$A$_{2n}$. It should be noted, that when \emph{n} is
increased to \emph{n}+1 the number of carbon atoms in the unit cell
is increased by eight atoms, due to the fact that eight carbon atoms
can form a full cycle with a sixfold ring in the projections along
the other \emph{b} or \emph{c}-axis. As such, the atom numbers of
the conventional unit cells of these families are equal to
8(\emph{m} + \emph{n}).

\begin{table}
\caption{DFT optimized lattice constants (\emph{a}, \emph{b}, and
\emph{c} in \AA\,) Wyckoff position(W.p.), bulk and shear moduli (B
and G in GPa), estimated Vickers hardness (H$_v$ in GPa) and
theoretical density ($\rho$, g/cm$^3$) for S, R and P allotropes.}
\begin{ruledtabular}
\begin{tabular}{ccccccccccccccccc}
Types & & W.p. & \emph{x}  & \emph{y}  & \emph{z}  & \\

\hline
S carbon              & \emph{a} = 4.7302     & 2\emph{m} & 0.1175 &0.0 & 0.6746 & $G$ = 457.7  \\
($P2/m$)         & \emph{b} = 2.4950          & 2\emph{m} & 0.5344 &0.0 & 0.3333 & $B$ = 412.6  \\
                       & \emph{c} = 4.0837    & 2\emph{n} & 0.1131 &0.5 & 0.8997 & $H_v$ = 78.3 \\
              & $\beta$ = 106.1$^\circ$     & 2\emph{n} & 0.4209 &0.5 & 0.1319 & $\rho$ = 3.44 \\
\hline
R carbon              & \emph{a} = 7.7886     & 4\emph{g} & 0.6731 & 0.9630 & 0.0 & $G$ = 462.4  \\
($Pbam$)         & \emph{b} = 4.7752          & 4\emph{g} & 0.8435 & 0.8087 & 0.0 & $B$ = 434.2  \\
                    & \emph{c} = 2.4958       & 4\emph{h} & 0.9546 & 0.8613 & 0.5 & $H_v$ = 75.0 \\
                                 &            & 4\emph{h} & 0.5704 & 0.8926 & 0.5 & $\rho$ = 3.45 \\
\hline
P carbon             & \emph{a} = 8.6650    & 4\emph{f} & 0.5357 & 0.25 & 0.4322& $G$ = 485.0  \\
($Pmmn$)         & \emph{b} = 2.4875        & 4\emph{f} & 0.2077 & 0.25 & 0.4348 & $B$ = 449.1  \\
                       & \emph{c} = 4.2160  & 4\emph{f} & 0.0414 & 0.25 &0.5625 & $H_v$ = 78.5 \\
                                 &          & 4\emph{f} & 0.7151 & 0.25 & 0.4343& $\rho$ = 3.51 \\
\end{tabular}
\end{ruledtabular}
\label{tab1}
\end{table}

\begin{figure}[b!]
\begin{center}
\includegraphics[width=0.4\textwidth]{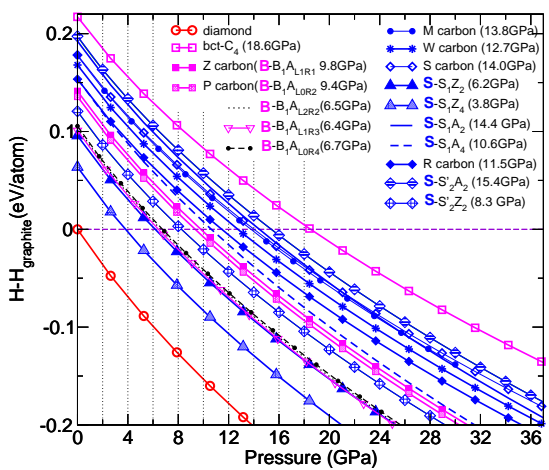}
\end{center}
\caption{(color online) Calculated enthalpy difference per atom with
respect to graphite for a series of carbon allotropes as a function
of pressure. The stable pressures
of these proposed allotropes compared to graphite are labeled.}\label{fig3}
\end{figure}

Figure 2a shows the 2D projection of bct-C$_4$, indicating the
simplest 4+8 rings pattern along the armchair orientation (direct
face-to-face opposed connection of two armchair chains). By
inserting the armchair chains into periodically separated 4+8 ring
pattern, a new structural family of B-B$_m$A$_{2n}$ can be
constructed. Here, B is the unit of the 4+8 rings composed of two
opposed armchair chains originated from bct-C$_4$ and A refers to
the armchair chain. For the sake of convenient comparisons, we
further split A$_{2n}$ into A$_{{Ln^{\prime}}{Rn^{\prime\prime}}}$
to distinguish the left- and right-armchair chain numbers of the B
unit. Clearly, $n^{\prime}+n^{\prime\prime}$ is equal to 2\emph{n},
indicating that the number of atoms is 8(\emph{m}+\emph{n}) in the
conventional unit cells. With \emph{m}=1 and \emph{n} =0, B-B$_1$
becomes bct-C$_4$. Furthermore, the addition of two armchair chains
with \emph{n}=1 separates the 4+8 ring pattern thus leading to two
new structures: the first one (Fig. 2b) corresponds to the Z carbon
(B-B$_1$A$_{L1R1}$) as proposed in Refs. \cite{Amsler,Selli,Zhao},
whereas we name the second one (Fig. 2c) \emph{P} carbon
(B-B$_{1}$A$_{L0R2}$). By adding four armchair chains, \emph{n}=2,
three alternative structures are formed, as depicted in Fig.
2\emph{d} (B-B$_{1}$A$_{L2R2}$), Fig. 2\emph{e}
(B-B$_{1}$A$_{L1R3}$) and Fig. 2\emph{f} (B-B$_{1}$A$_{L0R4}$).
Therefore, it is clear that all structures in the B-B$_m$A$_{2n}$
family are composed of 4+8+6 ring patterns. The introduction of
zigzag chains in the 4+8 ring patterns bct-C$_4$ will lead to the
occurrence of 4+8+5+7 ring complex structures. However, we do not
discuss these cases since all phases associated with 4+8+5+7 ring
complex (not shown here) are energetically much less stable than the
S- and B-families.

Figure 3 compares the relative thermodynamic stabilities of all
proposed allotropes. We find that, for both S and B families with
increasing zigzag or armchair units (namely, increasing the
6-membered rings in the unit cell) their structures become more
stable in energy and the transition pressure associated with the
structural transition decreases. This can be attributed to the
introduction of more ideal 6-membered rings which actually reduces
the total strain in system. If the 5+7 patterns (or 4+8 patterns)
are doubly coupled in a complementary mode, the energy of system can
be even lower. The complementary mode reduces the strain of system
compared to the cases in which a single 5+7 (or 4+8) pattern appears
in a non-coupled complementary one. Although Z and P phases share
the same composition of B-B$_1$A$_2$, P carbon is more stable in
energy than Z carbon (Fig. \ref{fig3}) because the 4+8 patterns in P
carbon are always coupled doubly in a complementary mode (Fig.
\ref{fig2}c), whereas in Z carbon they are separated by the
6-membered rings (Fig. \ref{fig2}b). In addition, in these families
as \emph{n} increases the energies of the structures approach more
closely that of diamond due to the increase in the number of ideal
6-membered rings. In this situation, the S (5+7 ring) and B (4+8
ring) patterns can be thus viewed as the defect or tilt grain
boundary in diamond. Following this viewpoint, the S-S$_m$Z$_{2n}$
and S-S$_m$A$_{2n}$ families can be interpreted as the combination
of diamond (6-membered ring) and S carbon whereas the
B-B$_m$A$_{2n}$ families show the combined character of diamond and
bct-C$_4$. Furthermore, from Fig. \ref{fig3} our selected phases P
carbon, B-B$_1$A$_{L0R4}$, B-B$_1$A$_{L1R3}$, S-S$_1$Z$_2$, and
S-S$_1$Z$_4$, are energetically more favorable than all previously
theoretically proposed structures including Z carbon. In particular,
even when their vibrational entropies and zero-point energies
derived from phonon densities of states are taken into account, the
relative stabilities for S-, R- and P-carbon allotropes remain
unchanged at least in the temperature range from 0 to 800 K at both
0 and 15 GPa.

\begin{figure}[b!]
\begin{center}
\includegraphics[width=0.40\textwidth]{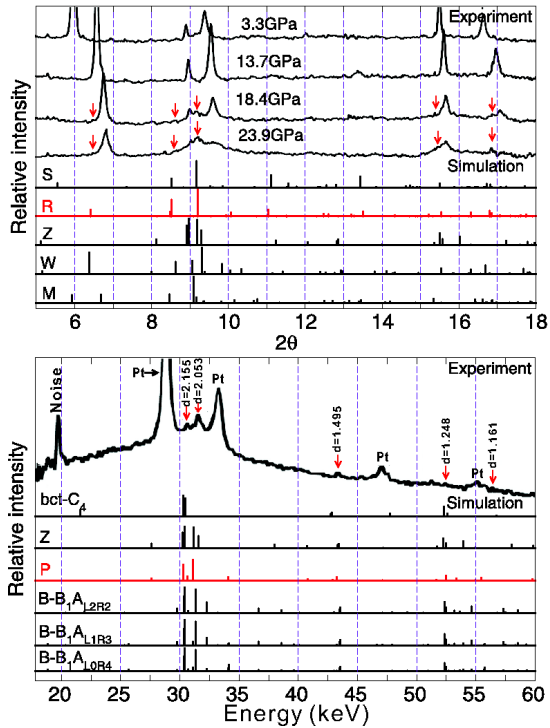}
\end{center}
\caption{(color online) (a) Simulated XRD patterns (with wavelength
of 0.3329 {\AA}) for selected phases at 18.4 GPa compared with
experimental results from Ref. \cite{Wendy03} for cold-compressed
graphite. (b) Simulated XRD patterns of the five allotropes in B
family at ambient pressure, in comparison with the date measured
from the quenchable phase recovered from cold-compressed CNTs
\cite{Wang2004}. The red arrows highlight five main experimental
peaks. \label{fig5}}
\end{figure}

Structural optimization revealed that R and P carbon crystallize in
the orthorhombic structure whereas S carbon has a monoclinic
structure (Figs. 1 and 2). The structural details of S, R, P and the
other nine new allotropes proposed here are listed in Tab. 1 and
Tab. S1 \cite{RefAdd}, respectively. The phonon dispersions in the
whole Brillouin zone for S, R and P carbon phases have been derived,
confirming their crystalline stabilities even at 0 GPa (Fig. S3
\cite{RefAdd}). In addition, our calculations uncovered that all
these allotropes exhibit large bulk and shear moduli (Table 1) which
are comparable to those of diamond. Utilizing our recently proposed
formula \cite{Chen1,Chen2}, the Vickers hardness of S, R and P
carbon are estimated to be 78.3, 75 and 78.5 GPa in Tab. 1,
respectively (details refer to Tab. S3 \cite{RefAdd}. This clearly
demonstrates that these phases are superhard in agreement with the
quasi-\emph{sp}$^3$-hybrid covalent bonding framework. The HSE
electronic band structures illustrate that S-, R-, and P-carbon are
all wide-gap transparent insulators with band gaps ranging between
3.9 and 5.5 eV (Fig. S3 \cite{RefAdd}). In particular, both S and R
carbon have the widest direct band gap at $\Gamma$ (about 5.5 eV)
among all known allotropes discussed here.

Although Z carbon was identified as a good candidate of cold
compressed graphite in the latest work \cite{Amsler}, this was
challenged by Zhao {\em et al.} \cite{Zhao} due to the problems for
matching experimental XRD patterns. They demonstrated that its
simulated patterns match neither the experimental peak at
16.8$^\circ$ and nor the gradually decreased peak density around
9.0$^\circ$ \cite{Zhao} for cold-compressed graphite (for details,
see the online supporting materials of Ref. \cite{Amsler}).
Importantly, we found that the simulated XRD patterns of R carbon
match much better the experimental data for five main peaks as
marked by arrows in Fig. \ref{fig5}(upper panel). The obtained
pressure of 11.5 GPa for the stability of R carbon is consistent
with the experimental value. Therefore, R carbon can be considered
as a highly likely candidate for cold-compressed graphite.

Furthermore, we found that P carbon is a better candidate to
interpret the experimental XRD pattern of cold-compressed CNTs than
Z carbon as claimed in Ref. \cite{Zhao}. From Fig. \ref{fig5} (lower
panel), for Z carbon and three other B-family phases
(B-B$_{1}$A$_{L2R2}$, B-B$_{1}$A$_{L1R3}$, B-B$_{1}$A$_{L0R4}$) the
simulated peaks at $d$ = 2.155 \AA\, are stronger (or at least
comparable) than the ones at 2.053 \AA\,. This notion is not
consistent with the reported XRD patterns \cite{Wang2004} which
clearly revealed the \emph{d}=2.053 \AA\ peak has the highest
intensity. This fact suggests that Z carbon may not be a good
candidate of cold-compressed CNTs. Nevertheless, it is exciting to
note that the simulated XRD pattern of P carbon shows a
significantly improved agreement with the experimental data. Its
five main peaks at $d$ = 2.167 \AA\,, 2.107 \AA\,, 1.510 \AA\,,
1.244 \AA\, and 1.179 \AA\, give satisfactory accordance in both
locations and intensities with experimental \emph{d} spacings of
2.155 \AA\,, 2.053 \AA\,, 1.495 \AA\,, 1.248 \AA\, and 1.161 \AA\,
respectively (see Fig. \ref{fig5}(lower panel)). Besides, its
theoretical bulk density ($\rho$= 3.51 g/cm$^3$) and bulk modulus
(\emph{B} = 449.1 GPa) are in nice agreement with the experimental
data ($\rho$ = 3.6$\pm$0.2 g/cm$^3$ and \emph{B} = 447 GPa), see
Tab. 1 and S2 \cite{RefAdd}.

In summary, R-carbon in the S-family and P-carbon in the B-family
have been found to best match with the experimental data of
cold-compressed graphite and CNTs, respectively. Moreover, for all
other proposed allotropes related to the S- and B-families, the
simulated XRD patterns also capture their main experimental
features. In general, with increasing numbers of zigzag or armchair
units, the main peaks remain essentially unchanged (see Fig. S4
\cite{RefAdd}), though their corresponding 2$\theta$ values are
slightly shifted. Considering that in real samples a pressure
gradient could exist \cite{Wendy03,Wang2004} and the tube diameters
of CNTs show a wide range from 1.8 to 5.1 nm \cite{Wang2004} which
may create the conditions of the formation of different phases
\cite{Saito05,Saito11}, it would be reasonable to expect that the
cold-compressed phase of graphite and CNTs can be interpreted as a
mixture of several of the proposed allotropes.

\textbf{Acknowledgements:} We gratefully acknowledge fruitful
discussions with Z. Z. Zhang, K.-M. Ho, and C. Franchini. This work
was supported by the ``Hundred Talents Project'' of Chinese Academy
of Sciences and from NSFC of China (Grand Number: 51074151) as well
as Beijing Supercomputing Center of CAS (including its Shenyang
branch). W.L.M. and S.W. were supported by EFree, an Energy Frontier
Research Center funded by the U.S. Department of Energy, Office of
Science, Office of Basic Energy Sciences under Award Number
DE-SG0001057.

\end{document}